\documentclass[fleqn,twoside]{article}
\usepackage{espcrc2}

\usepackage{graphicx}
\usepackage[figuresright]{rotating}

\newcommand{\AmS}{{\protect\the\textfont2
  A\kern-.1667em\lower.5ex\hbox{M}\kern-.125emS}}

\title{Neutrino physics from cosmological observations}

\author{Steen Hannestad\address{Department of Physics, University of Odense 
(SDU), Campusvej 55, DK-5230 Odense M, Denmark}}
       
\begin{document}

\def\lesssim{\raise0.3ex\hbox{$\;<$\kern-0.75em\raise-1.1ex\hbox{$\sim\;$}}}
\def\gtrsim{\raise0.3ex\hbox{$\;>$\kern-0.75em\raise-1.1ex\hbox{$\sim\;$}}}

\begin{abstract}
We review the current status of neutrino cosmology, focusing mainly
on the question of the absolute values of neutrino masses and the 
possibility of a cosmological neutrino lepton asymmetry.
\vspace{1pc}
\end{abstract}

\maketitle

\section{Neutrino masses}

The absolute value of neutrino masses are very difficult to measure
experimentally. On the other hand, mass differences between neutrino
mass eigenstates, $(m_1,m_2,m_3)$, 
can be measured in neutrino oscillation experiments.
Observations of atmospheric neutrinos suggest a squared mass 
difference of $\delta m^2 \simeq 3 \times 10^{-3}$ eV$^2$
\cite{Fukuda:2000np,Fornengo:2000sr}. While there are still
several viable solutions to the solar neutrino problem the so-called
large mixing angle solution gives by far the best fit with
$\delta m^2 \simeq 5 \times 10^{-5}$ eV$^2$ \cite{sno,Bahcall:2002hv}
(see also contributions by A. Hallin and A. Smirnov in the present
volume). 

In the simplest case where neutrino masses are
hierarchical these results suggest that $m_1 \sim 0$, $m_2 \sim 
\delta m_{\rm solar}$, and $m_3 \sim \delta m_{\rm atmospheric}$.
If the hierarchy is inverted 
\cite{Kostelecky:1993dm,Fuller:1995tz,Caldwell:1995vi,Bilenky:1996cb,King:2000ce,He:2002rv}
one instead finds
$m_3 \sim 0$, $m_2 \sim \delta m_{\rm atmospheric}$, and 
$m_1 \sim \delta m_{\rm atmospheric}$.
However, it is also possible that neutrino
masses are degenerate
\cite{Ioannisian:1994nx,Bamert:vc,Mohapatra:1994bg,Minakata:1996vs,Vissani:1997pa,Minakata:1997ja,Ellis:1999my,Casas:1999tp,Casas:1999ac,Ma:1999xq,Adhikari:2000as}, 
$m_1 \sim m_2 \sim m_3 \gg \delta m_{\rm atmospheric}$, 
in which case oscillation experiments are not
useful for determining the absolute mass scale.

Experiments which rely on kinematical effects of the neutrino mass
offer the strongest probe of this overall mass scale. Tritium decay
measurements have been able to put an upper limit on the electron
neutrino mass of 2.2 eV (95\% conf.) \cite{Bonn:tw}
(see also the contribution by Ch. Weinheimer in the present volume).
However, cosmology at present yields an even stronger limit which
is also based on the kinematics of neutrino mass.

Neutrinos decouple at a temperature of 1-2 MeV in the early universe,
shortly before electron-positron annihilation.
Therefore their temperature is lower than the photon temperature
by a factor $(4/11)^{1/3}$. This again means that the total neutrino
number density is related to the photon number density by
\begin{equation}
n_{\nu} = \frac{9}{11} n_\gamma
\end{equation}

Massive neutrinos with masses $m \gg T_0 \sim 2.4 \times 10^{-4}$ eV
are non-relativistic at present and therefore contribute to the
cosmological matter density \cite{Hannestad:1995rs,Dolgov:1997mb,Mangano:2001iu}
\begin{equation}
\Omega_\nu h^2 = \frac{\sum m_\nu}{92.5 \,\, {\rm eV}},
\end{equation}
calculated for a present day photon temperature $T_0 = 2.728$K. Here,
$\sum m_\nu = m_1+m_2+m_3$.
However, because they are so light
these neutrinos free stream on a scale of roughly 
$k \simeq 0.03 m_{\rm eV} \Omega_m^{1/2} \, h \,\, {\rm Mpc}^{-1}$
\cite{dzs,Doroshkevich:tq,Hu:1997mj}. 
Below this scale neutrino perturbations are completely erased and 
therefore the matter power spectrum is suppressed, roughly by
$\Delta P/P \sim -8 \Omega_\nu/\Omega_m$ \cite{Hu:1997mj}.

This power spectrum suppression allows for a determination of the
neutrino mass from measurements of the matter power spectrum on
large scales. This matter spectrum is related to the galaxy correlation
spectrum measured in large scale structure (LSS) surveys via the
bias parameter, $b^2 \equiv P_g(k)/P_m(k)$.
Such analyses have been performed several times before
\cite{Croft:1999mm,Fukugita:1999as}, most recently
using data from the 2dF galaxy survey \cite{Elgaroy:2002bi}. 
This investigation
finds an upper limit of 1.8-2.2 eV for the sum of neutrino masses.
However, this result is based on a relatively limited cosmological
parameter space. For the same data and an even more restricted
parameter space an upper limit of 1.5 eV was found \cite{Lewis:2002ah}.

It should also be noted that, although massive neutrinos have little
impact on the cosmic microwave background (CMB)
power spectrum, it is still necessary to include CMB data in any 
analysis in order to determine other cosmological parameters.

When calculating bounds on the neutrino mass from cosmological
observations great care must be taken, because if the analysis
is based on a too restricted parameter space, possible
parameter degeneracies cannot be studied and the bound on $m_\nu$
can become artificially strong.

\begin{figure}[h]
\begin{center}
\vspace*{-1cm}
\hspace*{-1cm}\includegraphics[width=20pc]{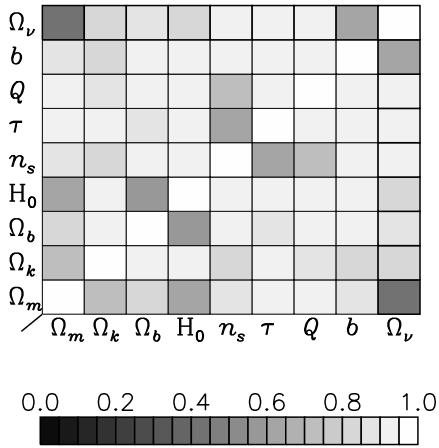}
\vspace*{-1cm}
\end{center}
\vspace*{-1.5cm}
\caption{Values of the parameter $r_{ij}$, defined in Eq.~(\ref{eq:rij}).}
\label{fig:1}
\end{figure}

In Fig. 1 the degeneracy between different cosmological
parameters is shown in form of the quantity
\begin{equation}
r_{ij} = \frac{\sigma_{j \,\, {\rm fixed}}(\theta_i)}{\sigma (\theta_i)}
\leq 1,
\label{eq:rij}
\end{equation}
i.e.\ the decrease in uncertainty of a measurement of parameter $i$
when parameter $j$ is fixed.
From this figure it can be seen that there are significant
degeneracies between $\Sigma m_\nu$ and other parameters, most
notably $b$, the bias parameter, and $\Omega_m$, the matter density.


In Ref.~\cite{han02} a full numerical likelihood analysis was
performed using a slightly restricted parameter space with 
the following free parameters:
$\Omega_m$, $\Omega_b$, $H_0$, $n_s$, $Q$, $b$, and $\tau$.
The analysis was further restricted to flat models, $\Omega_k=0$.
This has very little effect
on the analysis because there is little degeneracy between $m_\nu$ and
$\Omega_k$. In order to study the effect of the different priors
three different cases were calculated, the priors for which can be seen
in Table I. The BBN prior on $\Omega_b h^2$ comes from 
Ref.~\cite{Burles:2000zk}. The actual marginalization over parameters
other than $\Omega_\nu h^2$ was performed using a simulated annealing
procedure \cite{Hannestad:wx}.

Fig.~2 shows $\chi^2$ for the three different cases as a function of
the $m_\nu$.
The best fit $\chi^2$ values are 24.81, 25.66, and 25.71 for the
three different priors respectively. In comparison the number of
degrees of freedom are 34, 35, and 36, meaning that the fits are
compatible with expectations, roughly within the 68\% confidence
interval.

\begin{figure}[h]
\begin{center}
\hspace*{-1cm}\includegraphics[width=20pc]{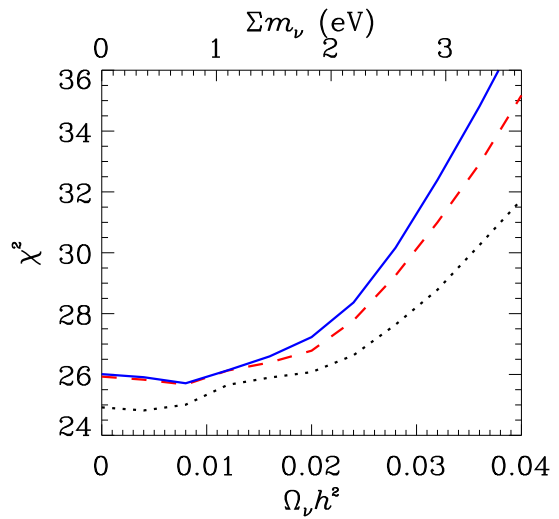}
\end{center}
\vspace*{-1cm}
\caption{$\chi^2$ as a function of $\Omega_\nu h^2$, plotted for
the three different priors. The dotted curve is for CMB+LSS, the
dashed for CMB+LSS+BBN+$H_0$, and the full curve for
CMB+LSS+BBN+$H_0$+SNIa.}
\label{fig:2}
\end{figure}

The 95\% confidence limit on $m_\nu$ was identified with the point
where $\Delta \chi^2 = 4$. These limits are shown in Table II.
For the most restrictive prior we find a 95\% confidence upper limit
of $\sum m_\nu \leq 2.47$ eV. This is compatible with the findings
of Ref.~\cite{Elgaroy:2002bi} 
who derived that $\sum m_\nu \lesssim 1.8-2.2$ eV
for a slightly more restrictive parameter space.

Based on the present analysis we consider 
$\sum m_\nu \leq 3$ eV (95\% conf.) a robust upper limit on the sum
of the neutrino masses. This corresponds roughly to the value found 
for the CMB+LSS data alone without any additional priors.
Even though this value is significantly higher than what is quoted
in Ref.~\cite{Elgaroy:2002bi}, 
it is still much more restrictive than the 
value $\sum m_\nu \leq 4.4$ eV \cite{WTZ} found from CMB and PSCz 
\cite{pscz} data. As is also
discussed in Ref.~\cite{Elgaroy:2002bi} 
the main reason for the improvement is
the much greater precision of the 2dF survey, compared to the PSCz
data \cite{pscz}.

\begin{table*}[htb]
\caption{The different priors on parameters other than $\Omega_\nu h^2$ 
used in the analysis of Ref.~\protect\cite{han02}.}
\label{table:1}
\newcommand{\m}{\hphantom{$-$}}
\newcommand{\cc}[1]{\multicolumn{1}{c}{#1}}
\renewcommand{\tabcolsep}{2pc} 
\renewcommand{\arraystretch}{1.2} 

\begin{tabular}{lccc}
\hline
Parameter & CMB + LSS & CMB + LSS & CMB + LSS + BBN  \\
&& + BBN + $H_0$ & + BBN + $H_0$ + SNIa \\
\hline
$\Omega_m$ & 0.1-1 & 0.1-1 & $0.28 \pm 0.14$ \\
$\Omega_b h^2$ & 0.008 - 0.040 & $0.020 \pm 0.002$ & $0.020 \pm 0.002$ \\
$h$ & 0.4-1.0 & $0.70 \pm 0.07$ & $0.70 \pm 0.07$ \\
$n$ & 0.66-1.34 & 0.66-1.34 & 0.66-1.34 \\
 $\tau$ & 0-1 & 0-1 & 0-1 \\
$Q$ & free & free & free \\
$b$ & free & free & free \\
\hline
\end{tabular}\\[2pt]
\end{table*}

\begin{table*}[htb]
\caption{Best fit $\chi^2$ and upper limits on $\sum m_{\nu,{\rm max}}$
for the three different priors.}
\label{table:2}
\newcommand{\m}{\hphantom{$-$}}
\newcommand{\cc}[1]{\multicolumn{1}{c}{#1}}
\renewcommand{\tabcolsep}{2pc} 
\renewcommand{\arraystretch}{1.2} 
\begin{tabular}{lcc}
\hline
prior type & best fit $\chi^2$ & $\sum m_{\nu,{\rm max}}$ (eV) (95\%) \\
\hline
CMB + LSS & 24.81 & 2.96 \\
CMB + LSS + BBN + $H_0$ &  25.66 & 2.65 \\
CMB + LSS + BBN + $H_0$ + SNIa & 25.71 & 2.47 \\
\hline
\end{tabular}\\[2pt]
\end{table*}

\section{Cosmological neutrino lepton asymmetry}

Cosmological observations, in addition to the information they 
yield on neutrino masses, also
provide a powerful means of determining the presence of a non-zero
cosmological neutrino lepton number or additional sterile neutrino 
species.

In addition to affecting structure formation neutrinos also 
contribute relativistic energy density in the early universe.
This has a profound effect on big bang nucleosynthesis as well
as the CMB formation.
The Friedmann equation, $H^2 = 8 \pi G \rho/3$, yields a relationship
between temperature and expansion rate.
The beta reactions which maintain equilibrium between neutrons
and protons in the early universe freeze out roughly when
$\Gamma/H \sim 1$. If relativistic energy density is added then
$H$ is larger for a given temperature and the beta reactions 
freeze out faster. The neutron to proton ratio is in equilibrium
given by $n/p \propto e^{-Q/T}$, with $Q = m_n - m_p = 1.293$ MeV,
so that if relativistic energy
density is added more neutrons survive. This in turn means that
more helium is formed, the dependence being roughly $Y_P \sim N_\nu^{1/6}$,
and indeed observations of the
primordial helium abundance can be used to 
constrain the amount of relativistic energy density.

The standard way to parameterise such energy density is
in equivalent number of neutrino species $N_\nu \equiv \rho_R/\rho_{\nu,0}$,
where $\rho_{\nu,0}$ is the energy density of a standard neutrino
species. At present the bound is roughly $N_\nu \lesssim 3.5-4$
\cite{Abazajian:2002bj,Lisi:1999ng}. 

At first sight this bound can be translated directly into a bound on
the neutrino lepton number as well, because a non-zero lepton number
yields additional energy density.
\begin{equation}
N_\nu = 3 + \frac{30}{7}\left(\frac{\mu}{\pi T}\right)^2
+  \frac{15}{7}\left(\frac{\mu}{\pi T}\right)^4,
\end{equation}
assuming that only one neutrino species has non-zero lepton number.

However, there is a fundamental difference in that the bound is
flavour sensitive. The electron flavour neutrinos enter directly
into the beta reactions, and an electron neutrino lepton number
therefore has a different influence on BBN than muon or tau
neutrino lepton numbers \cite{Kang:xa}.

In practise this means that a large positive chemical potential
in muon and tau neutrinos can be compensated by a small electron
neutrino chemical potential.
Of course such models are quite contrived, but it is highly
desirable with independent methods for determining the cosmological
neutrino lepton numbers.

It turns out that the CMB is at least in principle also an 
excellent probe of the relativistic energy density. The reason is
that an increase in the relativistic energy density delays matter
radiation equality, which in turn leads to an increase of the
so-called early integrated Sachs-Wolfe (ISW) effect. In the power
spectrum this shows up as an increase in power around the scale
of the particle horizon slightly after recombination, i.e\ around 
the scale of the first acoustic peak. This effect can be seen
in Fig. 3.

\begin{figure}[h]
\begin{center}
\hspace*{-1cm}\includegraphics[width=20pc]{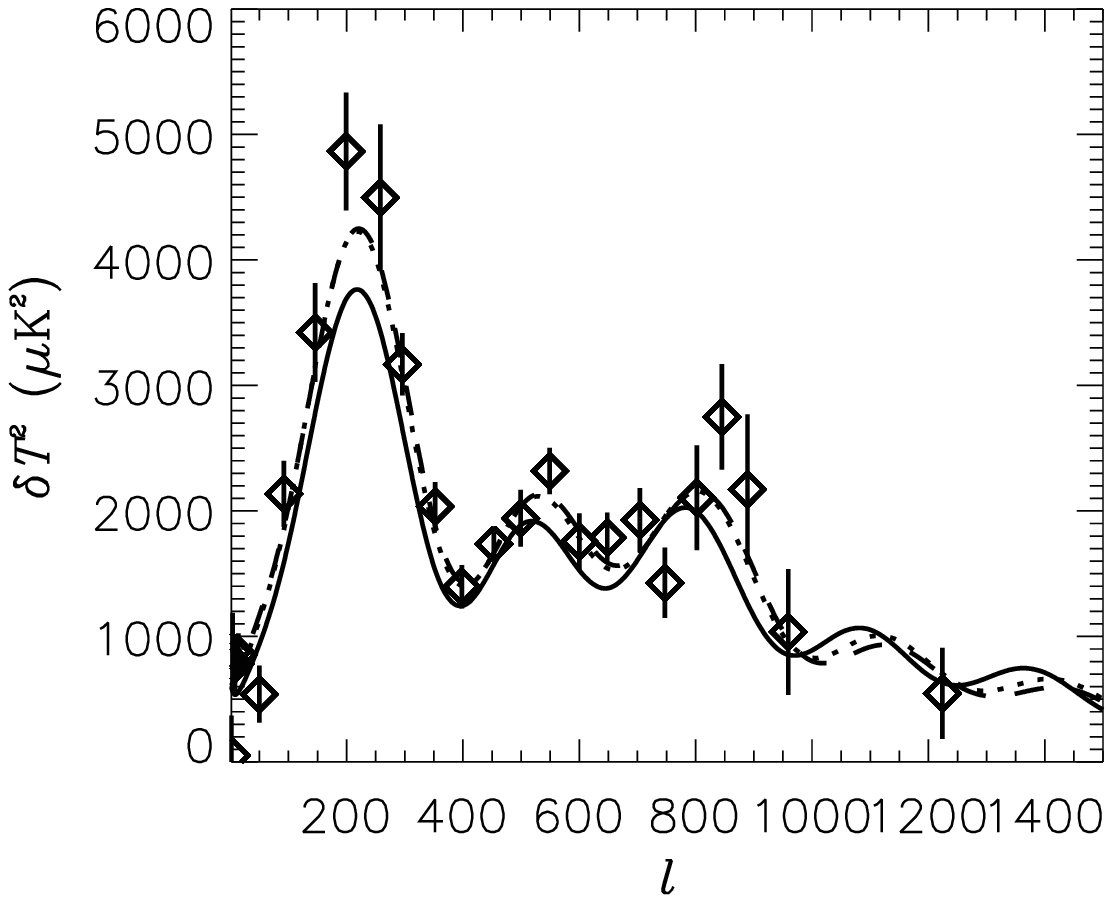}
\end{center}
\vspace*{-1cm}
\caption{CMB power spectra for the best fits to the cosmological data.
The full line is for $N_\nu = 0$, the dashed for $N_\nu = 7$, and 
the dotted for $N_\nu = 14$. The data points are from the compilation
by Wang, Tegmark and Zaldarriaga \cite{WTZ}.}
\label{fig:3}
\end{figure}

This effect has been used previously to constrain $N_\nu$ using
data from the best present day experiments 
\cite{Hannestad:2000hc,Hannestad:2001hn,Esposito:2000sv,Kneller:2001cd}, 
namely Boomerang \cite{boom},
Maxima \cite{max}, CBI \cite{cbi} and DASI \cite{dasi}. 
Unfortunately the data is not yet of sufficient
accuracy to yield constraints anywhere near as strong as BBN. However,
on the other hand they do not suffer from the same problems of being
flavour sensitive. It is therefore possible to combine BBN
and CMB constraints to yield a non-trivial bound on neutrino
lepton chemical potentials \cite{Hansen:2001hi}.
From a combination of CMB data with the PSCz data a bound of 
$N_\nu = 6^8_{4.5}$ (95\%) has been obtained \cite{Hannestad:2001hn}
(see also Table 3).
Interestingly, this has provided the first detection of the 
cosmic neutrino background around a redshift of 1000, and the first
measurement independent of big bang nucleosynthesis.

It should also be noted that the flavour dependence of the BBN bound
relies on the assumption that there is no mixing between different 
flavours during BBN. However, if the solar LMA solution turns out to
be correct it is almost inevitable that there is significant equilibration
between $\nu_e$ and $\nu_\mu$ prior to BBN. It has been shown that
this significantly strengthens the bound because a large $\nu_\nu$ or
$\nu_\tau$ lepton number will leak into the electron sector before BBN
\cite{lunardini,Pastor:2001iu,Dolgov:2002ab,Abazajian:2002qx,Wong:2002fa}.

\section{Discussion}

Cosmology offers an interesting probe of neutrino physics which 
is complementary to terrestrial experiments.

Observations of the large scale structure power spectrum has already 
given bounds on the absolute value of neutrino masses which
are comparable to, or stronger than the present bound from
tritium endpoint measurements. For the present data an upper
limit to the neutrino mass of $\sum m_\nu \lesssim 2.5-3$ eV can be
derived. This can be compared to the present bound from the Mainz 
experiment of $m_{\nu_e} = \sum_i  \left(|U_{ei}|^2 m_i^2\right)^{1/2} 
= 2.2$ eV.

However, it should be noted that an even stronger upper bound
can be put on neutrino masses if they are Majorana particles.
In that case neutrinoless double beta decay is possible
because lepton number is not a conserved quantity. The
non-observation of such events has led to the bound
\begin{equation}
m_{ee} = \sum_j U^2_{ej} m_{\nu_j} < 0.27 \,\, {\rm eV},
\end{equation}
where $U$ is the neutrino mixing matrix \cite{klapdor}. 
From observations of neutrinoless double beta decay it has indeed
been claimed that positive evidence for non-zero neutrino masses
has been obtained, with a favoured value in the range 0.11-0.56 eV
\cite{Klapdor-Kleingrothaus:2001ke}.
However, this claim is highly controversial and has been refuted
by a number of other authors \cite{Aalseth:2002dt}. 
At present it therefore seems safest
to regard neutrinoless double beta decay experiments as yielding
only an upper limit on $N_\nu$.

In the coming years the large scale structure power spectrum will
be measured even more accurately by the Sloan Digital Sky Survey,
and at the same time the CMB anisotropy will be probed to great
precision by the MAP and Planck satellites. By combining these
measurements it was estimated by Hu, Eisenstein
and Tegmark that a sensitivity of about 0.3 
eV could be reached \cite{Hu:1997mj}.
Currently an upgrade of the Mainz experiment by an order of magnitude,
the KATRIN experiment,
is planned. Such an experiment should take the limit on the electron neutrino
mass down to about 0.2 eV.
The prospects for measuring a neutrino mass of the order 0.1 eV,
as suggested by oscillation experiments is therefore almost within reach.

Another cosmological probe of the neutrino mass is the so-called
Z-burst scenario for ultrahigh energy cosmic rays 
\cite{Weiler:1997sh,Fargion:1997ft}. 
Neutrinos are
not subject to the GZK cut-off which applies to protons
\cite{Greisen:1966jv,Zatsepin:1966jv}. Therefore it is
in principle possible that the primary particles for super GZK
cosmic rays are neutrinos. One possibility which has been explored
is that the neutrino-nucleon cross-section increases drastically
at high CM energies, for instance due to the presence of large 
extra dimensions.
The other possibility is that neutrinos have rest mass in the 
eV range. In that case high energy neutrinos can annihilate
on cosmic background neutrinos with a large cross section if the 
CM energy is close to the Z-resonance, corresponding to a primary
neutrino energy of $E_\nu \simeq 4 \times 10^{21} m_{\rm eV}^{-1} \, {\rm eV}$.
This annihilation would produce high energy protons which could then
act as primaries for the observed high energy cosmic rays.
The observed ultrahigh energy cosmic ray flux can be explained 
if the heaviest neutrino has a mass larger than $\sim 0.1$ eV. 
Therefore, if the Z-burst scenario turns out to be correct,
it is in principle possible to measure a neutrino mass in this range
\cite{Pas:2001nd,Fodor:2001qy,Ringwald:2001mx}.

With regards to the neutrino relativistic energy density, the present
bound from BBN is roughly $N_\nu \lesssim 3.5-4$. The prospects for
improving this bound in the future do not seem too bright. The present
bound is most likely dominated by systematic effects in the measurement
of helium abundances. BBN is also a very powerful probe of a possible
neutrino lepton asymmetry, particularly if, as is indicated by 
data from SNO and Super-Kamiokande \cite{sno,Bahcall:2002hv}, 
the LMA mixing solution is correct.
In this case, an upper bound on the lepton asymmetry for any flavour
is roughly $\mu_{\nu_i}/T \lesssim 0.07$ \cite{Dolgov:2002ab}.

The combination of CMB and large scale structure data can also be used for
constraining the neutrino relativistic energy density. At present the
CMB data are not of sufficient accuracy to yield a bound which is
competitive with that from BBN, the present bound being 
$N_\nu = 6^8_{4.5}$ (95\%) (see also Table 3).
However, this bound applies to a very different epoch and therefore puts
significant constraints on possible entropy production after BBN
but before CMB formation.

\begin{table*}[htb]
\caption{Best fit values and $2\sigma$ (95\%) limits on $N_\nu$ for
different priors and two different data sets. Both priors
and data sets are discussed in Ref.~\protect\cite{Hannestad:2001hn}}
\label{table:3}
\newcommand{\m}{\hphantom{$-$}}
\newcommand{\cc}[1]{\multicolumn{1}{c}{#1}}
\renewcommand{\tabcolsep}{2pc} 
\renewcommand{\arraystretch}{1.2} 
\begin{tabular}{lcc}
\hline
prior type & WTZ & COBE+Boomerang \\
\hline
CMB only & $8^{+11}_{-8}$ & $7^{+17}_{-7}$ \\
CMB + BBN + $H_0$ & $8^{+9.5}_{-7}$ & $4^{+13}_{-4}$ \\
CMB + BBN + $H_0$ + LSS & $6^{+8}_{-4.5}$ & $9^{+8}_{-6.5}$  \\
\hline
\end{tabular}
\end{table*}                                                                     
In the near future a much more accurate determination of $N_\nu$ from
CMB measurements will become possible thanks to the satellites
MAP and Planck. It was estimated by Lopez et al. 
\cite{Lopez:1999aq} that it would
be possible to measure $\Delta N_\nu \sim 0.04$ using Planck data.
However, this is probably overly optimistic and a more reasonable
estimate seems to be $\Delta N_\nu \sim 0.1-0.2$ \cite{Bowen:2001in}.

This will also allow for a possible detection of sterile neutrinos
mixing with ordinary neutrinos in the early universe over a wide
range of parameter space \cite{Hannestad:1998zg}.

\end{document}